%%%%%%%%%%%%%%%%%%%%%%%%%%%%%%%%%%%%%%%%%%%%%%%%%%%%%%%%%%%%%%%%%%%%
%  TeX Definitions                                                 %
%%%%%%%%%%%%%%%%%%%%%%%%%%%%%%%%%%%%%%%%%%%%%%%%%%%%%%%%%%%%%%%%%%%%

\newif\iffigs\figstrue
% Uncomment the next line if you do not want the figures
%\figsfalse

%
% the following is to use blackboard bold fonts --
\let\useblackboard=\iftrue
%
% activate this if you don't have them.
\let\useblackboard=\iffalse
%
% You might also need to remove this line.
%\newfam\black
%

\input harvmac.tex
\input epsf
\newcount\figno
\figno=0
\def\fig#1#2#3{
\par\begingroup\parindent=0pt\leftskip=1cm\rightskip=1cm\parindent=0pt
\baselineskip=11pt
\global\advance\figno by 1
\midinsert
\epsfxsize=#3
\centerline{\epsfbox{#2}}
\vskip 12pt
{\bf Fig.\ \the\figno: } #1\par
\endinsert\endgroup\par
}
\def\figlabel#1{\xdef#1{\the\figno}}
\def\encadremath#1{\vbox{\hrule\hbox{\vrule\kern8pt\vbox{\kern8pt
\hbox{$\displaystyle #1$}\kern8pt}
\kern8pt\vrule}\hrule}}
\overfullrule=0pt

\def\Title#1#2{\rightline{#1}
\ifx\answ\bigans\nopagenumbers\pageno0\vskip1in%
\baselineskip 15pt plus 1pt minus 1pt
\else%\special{papersize=11in,8.5in}%
\def\listrefs{\footatend\vskip 1in\immediate\closeout\rfile\writestoppt
\baselineskip=14pt\centerline{{\bf References}}\bigskip{\frenchspacing%
\parindent=20pt\escapechar=` \input
refs.tmp\vfill\eject}\nonfrenchspacing}
\pageno1\vskip.8in\fi \centerline{\titlefont #2}\vskip .5in}

\ifx\answ\bigans\def\tcbreak#1{}\else\def\tcbreak#1{\cr&{#1}}\fi
\useblackboard
\message{If you do not have msbm (blackboard bold) fonts,}
\message{change the option at the top of the tex file.}
\font\blackboard=msbm10 %scaled \magstep1
\font\blackboards=msbm7
\font\blackboardss=msbm5
%\newfam\black
\textfont\black=\blackboard
\scriptfont\black=\blackboards
\scriptscriptfont\black=\blackboardss

\else

\fi
% *************************************
%

\def\half{{1\over 2}}

%%%%%%%%%%%%%%%%%%%%%%%%%%%%%%%%%%%%%%%%%%%%%%%%%%%%%%%%%%%%%%%%%%%%%%%%%%%%
%                    F I G U R E S                                         %
%%%%%%%%%%%%%%%%%%%%%%%%%%%%%%%%%%%%%%%%%%%%%%%%%%%%%%%%%%%%%%%%%%%%%%%%%%%%
%\figsfalse

\iffigs
  \input epsf
\else
  \message{No figures will be included.  See TeX file for more
information.}
\fi

\input epsf

                   % N=? SUSY
\def\lbr{{\lbrack}}                             % [
\def\rbr{{\rbrack}}                             % ]

                              % wedge product
                              % Wilson lines
                              % inverse
                           % O(x)
               	% Real numbers
               	% Complex numbers
%%% \def\MR#1{{{\bf R}^{#1}}}               % Real numbers
%%% \def\MC#1{{{\bf C}^{#1}}}               % Complex numbers
               % Real numbers
               % Complex numbers
               % Circle, sphere,...
               	% disk, ball,...
              	 % Torus
             	 % CP
              	 % Ruled surface F_n
             	% Patch
                    	% line-bundle
              	% derivative
                 	% Left large bracket
                	% Right large bracket
              	% SL(*,Z)
                             	% identity matrix
      	% commutator
               	% anti-commutator
           	% expectation value
    	% expectation value of trace
      	% trace
            		% trace
            	% Trace
            	% trace in a rep
            	% Trace in a rep
                      		% representation
                  		% Imaginary
                  		% Imaginary
%\def\widebar#1{{\bar{#1}}}                    	% Wide bar
                    	% Wide bar
                 	% Pauli matrix
                      	% correction O()
                     		% Normal bundle
                     	 	% Hodge star
                         		% sign
\def\hepth#1{{\it hep-th/{#1}}}
\def\frac#1#2{{{{#1}}\over {{#2}}}}           	% {} over {}
\def\e8{E_8 \times E_8}                       	%E_8 times E_8
    	%S^1 over Z_2
%%%%%%%%%%%%%%%%%%%%%%%%%%%%%%%%%%%%%%%%%%%%%%%%%%%%%%%%%%%%%%%%%%%%%%%%%%%%
%                    Greek                                                 %
%%%%%%%%%%%%%%%%%%%%%%%%%%%%%%%%%%%%%%%%%%%%%%%%%%%%%%%%%%%%%%%%%%%%%%%%%%%%

%%% ------------------------- CUT HERE ---------------------------------%

%%%%%%%%%%%%%%%%%%%%%%%%%%%%%%%%%%%%%%%%%%%%%%%%%%%%%%%%%%%%%%%%%%%%%%%%%%%%
%                    TITLE PAGE                                            %
%%%%%%%%%%%%%%%%%%%%%%%%%%%%%%%%%%%%%%%%%%%%%%%%%%%%%%%%%%%%%%%%%%%%%%%%%%%%

%
\Title{ \vbox{\baselineskip12pt\hbox{hep-th/9803031, PUPT-1771}}}
{\vbox{
\centerline{Noncommutative Geometry from D0-branes}
\centerline{in a Background B-field}}}
\centerline{Yeuk-Kwan E. Cheung\footnote{$^1$}{email: cheung@viper.princeton.edu.}  and Morten Krogh\footnote{$^2$}{email: krogh@princeton.edu}}
\smallskip
\smallskip
\centerline{Department of Physics, Jadwin Hall}
\centerline{Princeton University}
\centerline{Princeton, NJ 08544-0708, USA}
%%%
\bigskip
\bigskip
\noindent
We study D0-branes in type IIA on $T^2$ with a background B-field turned on.  
We calculate explicitly how the background B-field modifies the D0-brane action.   
The effect of the B-field is to replace ordinary multiplication with a noncommutative product.   
This enables us to find the matrix model for M-theory on $T^2$ with a background 3-form 
potential along the torus and the lightlike circle.
This matrix model is exactly the non-local 2+1 dim SYM theory on a dual $T^2$ 
 proposed by Connes, Douglas and Schwarz. 
We calculate the radii and the gauge coupling for the SYM on the dual $T^2$ for all 
choices of longitudinal momentum  and membrane wrapping number on the 
$T^2$.

\Date{March 1998}

%%% ------------------------- CUT HERE ---------------------------------%

%%%%%%%%%%%%%%%%%%%%%%%%%%%%%%%%%%%%%%%%%%%%%%%%%%%%%%%%%%%%%%%%%%%%
%  B I B L I O G R A P H Y                                         %
%%%%%%%%%%%%%%%%%%%%%%%%%%%%%%%%%%%%%%%%%%%%%%%%%%%%%%%%%%%%%%%%%%%%

\lref\Connes{Alain Connes, Michael R. Douglas and Albert Schwarz,
  {``Noncommutative Geometry and Matrix Theory: Compactification on 
     Tori,''} \hepth{9711162.}}

\lref\DougHull{Michael R. Douglas and Chris Hull,
  {\it ``D-branes and the Noncommutative Torus,''}
  \hepth{9711165.}}

\lref\Sen{Ashoke Sen,
  {\it ``D0 Branes on $T^n$ and Matrix Theory,''} \hepth{9709220.}}

\lref\BFSS{T. Banks, W. Fischler, S.H. Shenker and L. Susskind,
  {\it ``M Theory As A Matrix Model: A Conjecture'',
   Phys. Rev. D55(1997)5112-5128} \hepth{9610043.}}

\lref\Seiwhy{N. Seiberg,
  {\it ``Why is the Matrix Model Correct?'',
  Phys.Rev.Lett. 79 (1997) 3577-3580} \hepth{9710009.}}

\lref\Wati{Washington Taylor,
  {\it ``D-brane field theory on compact spaces'',
  Phys.Lett. B394 (1997) 283-287} \hepth{9611042.}}

\lref\WOS{Ori J. Ganor, Sanjaye Ramgoolam and Washington Taylor,
  {\it ``Branes,Fluxes and Duality in M(atrix)-Theory'',
  Nucl.Phys.B492 (1997) 191-204} \hepth{9611202.}}

\lref\Li{Miao Li,
  {\it ``Comments on Supersymmetric Yang-Mills Theory on
         a Noncommutative Torus'',} \hepth{9802052.}}

\lref\HWW{Pei-Ming Ho,Yi-Yen Wu and Yong-Shi Wu,
  {\it ``Towards a Noncommutative Geometric Approach to 
         Matrix Compactification'',} \hepth{9712201.}}

\lref\Ming{Pei-Ming Ho and Yong-Shi Wu,
  {\it ``Noncommutative Gauge Theories in Matrix Theory,''}
   \hepth{9801147.}}

\lref\Casal{R Casalbuoni,
  {\it ``Algebraic treatment of compactification on 
         noncommutative tori'',} \hepth{9801170.}}

\lref\Halpern{M. Claudson and M. Halpern,
  {\it Nucl.Phys. B250 (1985) 689.}}

\lref\Flume{R. Flume,
  {\it Ann. of Phys. 164 (1985) 189.}}

\lref\BRR{M. Baake, P. Reinicke and V. Rittenberg,
  {\it J.Math. Phys. 26 (1985) 1070.}}

\lref\Witp{Edward Witten,
  {\it ``Bound States of Strings and p-Branes'',
  Nucl.Phys.B460 (1996) 335} \hepth{9510135.}}

\lref\Giveon{A. Giveon, M. Porrati and E. Rabinovici,
  {\it ``Target Space Duality in String Theory'',
  Phys.Rept. 244 (1994) 77-202} \hepth{9401139.}}

%%% ------------------------- CUT HERE ---------------------------------%

% ===================================================================== %
% Introduction
% ===================================================================== %
\newsec{Introduction}

Last fall Connes, Douglas and Schwarz \Connes\ made a very interesting 
proposal relating the matrix theory of M-theory 
on $T^2$ with a background three form potential, $C_{-12}$, to 
a gauge theory on a noncommutative torus. Shortly 
after Douglas and Hull justified this claim by relating 
these theories to a theory on a D-string \DougHull. 
They also mentioned that this could be seen in the 
original 0-brane picture.
	
The purpose of the present paper is to precisely 
incorporate the background B-field in the dynamics 
of 0-branes. In this way we will obtain the 
matrix model for M-theory on $T^2$ with $C_{-12} 
\neq 0$. It confirms the claims made by Connes,Douglas 
and Schwarz. The theory is a SYM theory on 
a dual torus with a modified interaction 
depending on the B-field. The theory 
contains higher derivative terms of arbitrarily 
high power and is thus non-local. We 
calculate the radii of the dual torus and the 
gauge coupling constant. We get a noncommutative 
gauge theory for any choice of longitudinal momentum 
and number of membranes wrapped 
around the $T^2$. The radii and gauge coupling   
depend on these numbers.

Aspects of the connection between compactifications 
of M-theory and noncommutative 
geometry has, among others, been worked out in 
\refs{\HWW, \Ming, \Casal}.

%%% ------------------------- CUT HERE ---------------------------------%

% ===================================================================== %
% Section (2): Zero-branes on T^2 with background B-field
% ===================================================================== %
\newsec{Zero-branes on $T^2$ with background B-field}

Let us consider M-theory on $T^2$ with radii, $R_1$, $R_2$. 
The torus is taken to be rectangular for simplicity. 
Making the torus oblique does not introduce anything interesting.
 We are interested in the matrix 
description of this
theory with a background $C$-field, ${C_{-12}\neq 0}$.  Here - denotes the lightlike circle
and $1, 2$ the directions along the torus.  Let the Plank mass be M and the radius 
of the lightlike circle be $R$.  Following \refs{\Sen,\Seiwhy}, we take this to mean a limit of 
spatial compactifications.  We also perform a rescaling to keep the interesting energies
finite.  The upshot is that we consider M theory on $T^2 \times S^1$ with Plank mass $\tilde M$
and radii $\tilde R_1$, $\tilde R_2$, $\tilde R$ in the limit ${\tilde R \rightarrow 0}$ with
\eqn\rescaling{\eqalign{
\tilde M^2 \tilde R &= M^2 R \cr
\tilde M \tilde R_i &= MR_i \qquad i=1,2. 
}}
This turns into type IIA on $T^2$ with string mass $m_s$, coupling $\lambda$ and radii
$r_1$, $r_2$ given by
\eqn\IIAscaling{\eqalign{
{m_s}^2 &= \tilde M^3 \tilde R \cr
\lambda &=(\tilde M\tilde R)^{3 \over 2}\cr
r_i&=\tilde R_i \qquad i=1,2. 
}}
Furthermore there is a flux of the $B_{ij}$ field through the torus.  We call the flux $B$:
\eqn\Bflux{
	B=R R_1 R_2 C_{-12}. 
}
We are interested in the sector of theory labelled by two integers, namely the number of 
D0-branes, $N_0$, and the number of D2-branes, $N_2$, wrapped on $T^2$.  In this section we
will solely be interested in the case $N_2=0$.  In the next section we will treat the general 
case.  Let us for the moment set $N_0$=1.  This makes us avoid some essentially irrevalent 
indices.  The generalization to general $N_0$ is straightforward.

The method for dealing with this situation has been developed in \refs{\Wati, \WOS}.  We work with 
the covering space of $T^2$, namely ${\bf R^2}$ and place 0-branes in a lattice (Figure 1).

\fig{0-branes on the covering space,$\bf R^2$ }{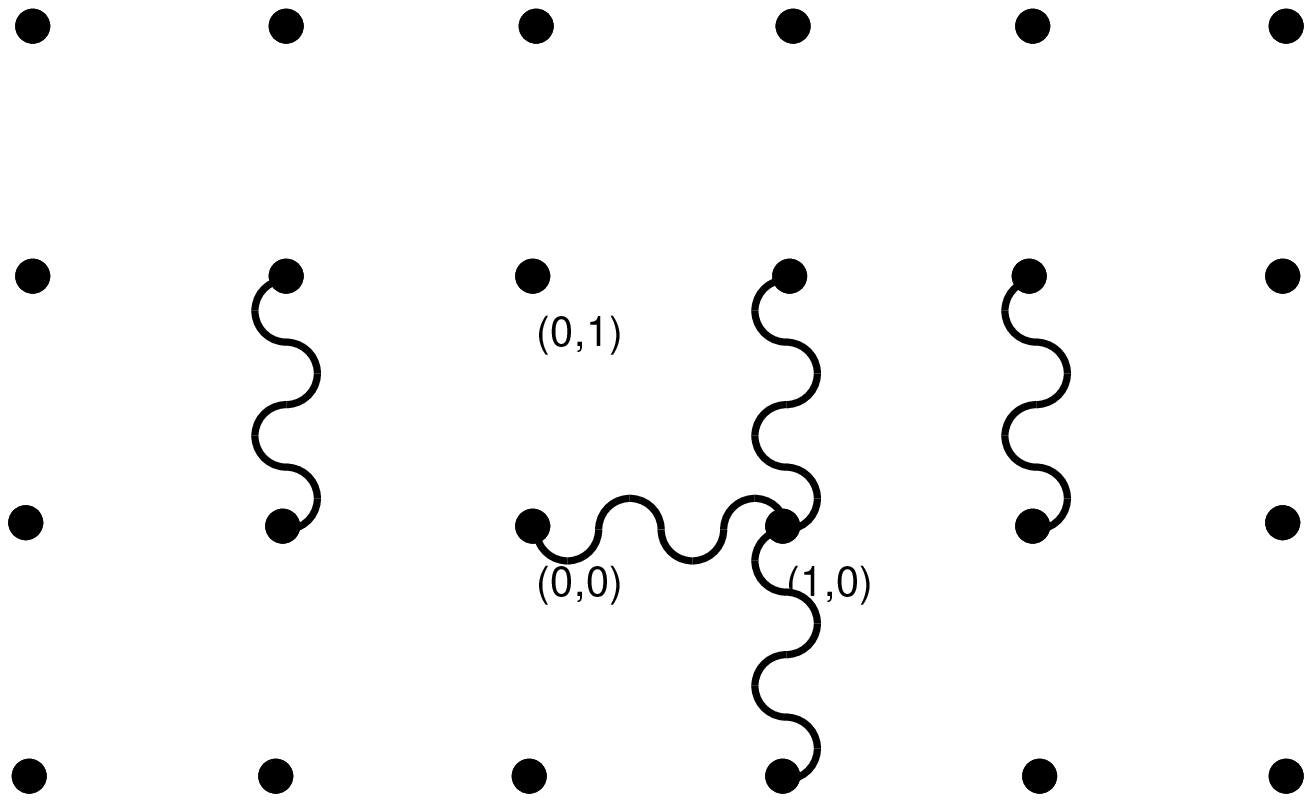}{4truein}

Let us label the 0-branes $(a, b)$  where $a$, $b$ are integers.  The open strings obey Dirichlet boundary
condition on the 0-branes.  This is the point where we need $N_2=0$.  If there had been D2-branes,
 the 0-branes would have been dispersed as fluxes inside the 2-branes and the open strings would
have Dirichlet boundary conditions on the 2-branes and the above picture does not apply.

The fields in the theory come from quantizing the open strings and calculating their interactions.
In the limit we are taking, ${m_s\rightarrow \infty}$, only the lowest modes survive and when 
$B=0$ the theory becomes a $SYM$ quantum mechanics \refs{\Halpern, \Flume, \BRR, \Witp, \BFSS}.  
The gauge group is $"U(\infty)"$ since there are infinitely many 0-branes.  To be more precise let us
define a Hilbert space on which the fields will be operators.  There is a basis vector for each 
0-brane, i.e. the basis is $|a,b>$ where $a, b \in {\bf Z}$.  Let $\phi$ be any field in theory, then 
the matrix element $\phi_{a_1b_1,a_2b_2}$ has the interpretation as a field which annihilates a 
state of an open string starting at $a_1b_1$ and ending at $a_2b_2$, see 
figure 2.
\fig{String starting at 0-brane $(a_1,b_1)$ and ending at 
     0-brane $(a_2,b_2)$} {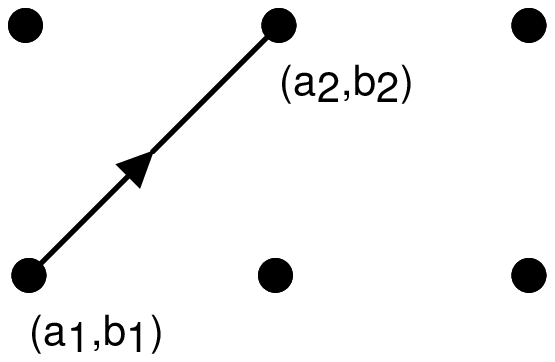} {2truein}
The fields of the theory are
\item{1.} bosons:$ \; X^i \qquad i=1,\ldots,8$ 
\item{2.} fermions:$ \; \Psi_{\alpha} \qquad \alpha=1,\ldots,16$ 
\item{3.} The gauge field: $A_0$.

The fields are constrained to obey the following conditions:
\eqn\constraints{\eqalign{
U_i^{-1}\;X^a\; U_i\; &= X^a + 2\pi r_a \delta^{ai}\;\; i=1,2; \cr
U_i^{-1}\;\Psi^{\alpha}\;U_i\;&=\; \Psi^{\alpha}, \cr
U_i^{-1}A_0 U_i &= A_0;
}}
where $U_i$ are translation operators on the states in the Hilbert space:
\eqn\defineu{\eqalign{
U_1|a,b>&=\;|a+1, b> \cr
U_2|a,b>&=\;|a,b+1>.
}}
The gauge field $A^0$ can be gauged away, and we will work in the gauge $A^0=0$. 
When $B=0$ the action is
\eqn\Obraneaction{\eqalign{
L= {m_s\over 2\lambda} \bf{Tr} \lbr
& \dot{X^a} \dot{X^a} + {m_s^4 \over (2\pi)^2} \sum_{a<b} {\lbr{X^a,X^b}\rbr}^2 \cr
 +& {m_s^2 \over {2\pi}} \Psi^T i \dot{\Psi} - {m_s^4 \over (2\pi)^2} \Psi^T
  \Gamma_a \lbr {X^a,\Psi} \rbr \rbr .
}}
What about $B\neq0$?  We will now show how to incorporate $B$-dependence in the action.
The two-form $B_{ij}$ couples to the worldsheet through the interaction $\int_{W.S.} B_{ij}$,
i.e. the $B$-field is pulled back to the worldsheet and integrated.  Let us look at the example shown 
in Figure 3 below.
\fig{The interaction between these three strings give rise 
      to a cubic vertex.} {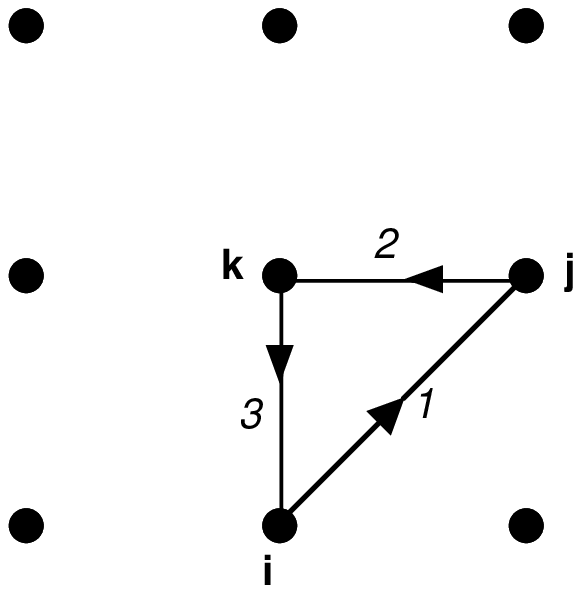} {2truein}
This interaction is represented, in the case $B=0$, by a term:
\eqn\threephi{ \kappa
\phi^{(3)}_{ik} \phi^{(2)}_{kj} \phi^{(1)}_{ji}
}
where $\kappa$ is a constant.  This term could, for instance, annihilate string 1 and 2
and create 3 with opposite orientation.  The worldsheet would look 
as shown in figure 4.
\fig{The worldsheet for a cubic vertex}{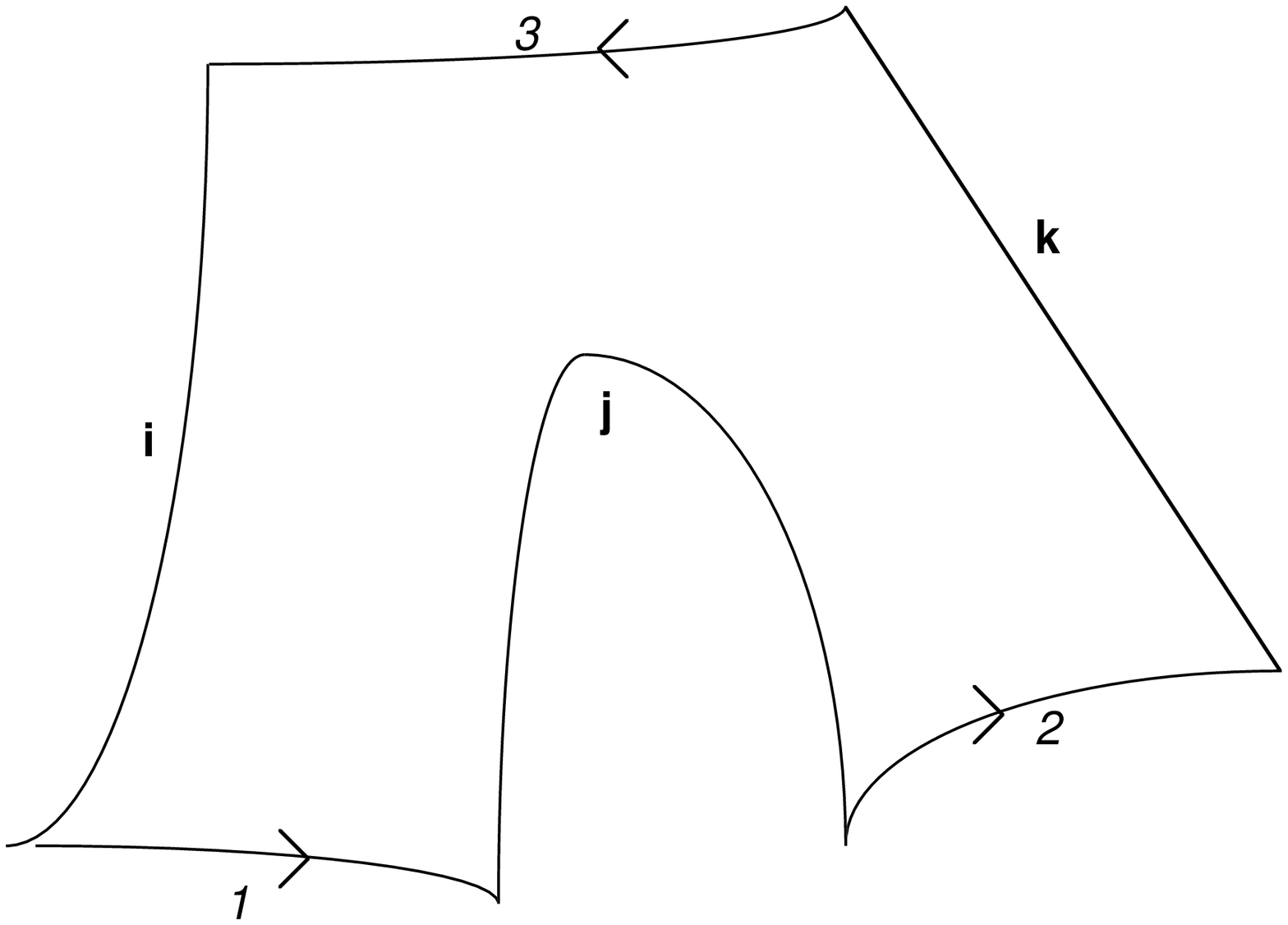}{4truein}
To calculate $\int_{W.S.} B_{ij}$ we just need the projection into the plane of the torus, since 
this is the only direction in which $B_{ij}\neq0$.  This projection is exactly given by the area 
between the three strings in Figure 3.  The important point is that $B_{ij}$ is closed so
$\int B_{ij}$ only depends on the homotopy type of the worldsheet imbedding and is 
insensitive to the finer details of how the interaction takes place.  For the example in Figure 3,
$\int_{W.S.} B_{ij} =\half B$, where we remark that $B$ was defined to be the flux through the 
torus.  This means that the interaction eq.\threephi\ now is replaced with
\eqn\Bthreephi{
e^{i\half B} \kappa \phi^{(3)}_{ik} \phi^{(2)}_{kj} \phi^{(1)}_{ji}.
}
It is now a straightforward exercise to figure out what happens to a general interaction 
between the fields:
\eqn\kphi{
	\phi^{(k)}_{a_1b_1,a_kb_k} \ldots \phi^{(2)}_{a_3b_3,a_2b_2} 
\phi^{(1)}_{a_2b_2,a_1b_1}.
}
We have to find the integral of the $B$-field through the polygon shown in Figure 5.
\fig{Generic form of a vertex with k strings} {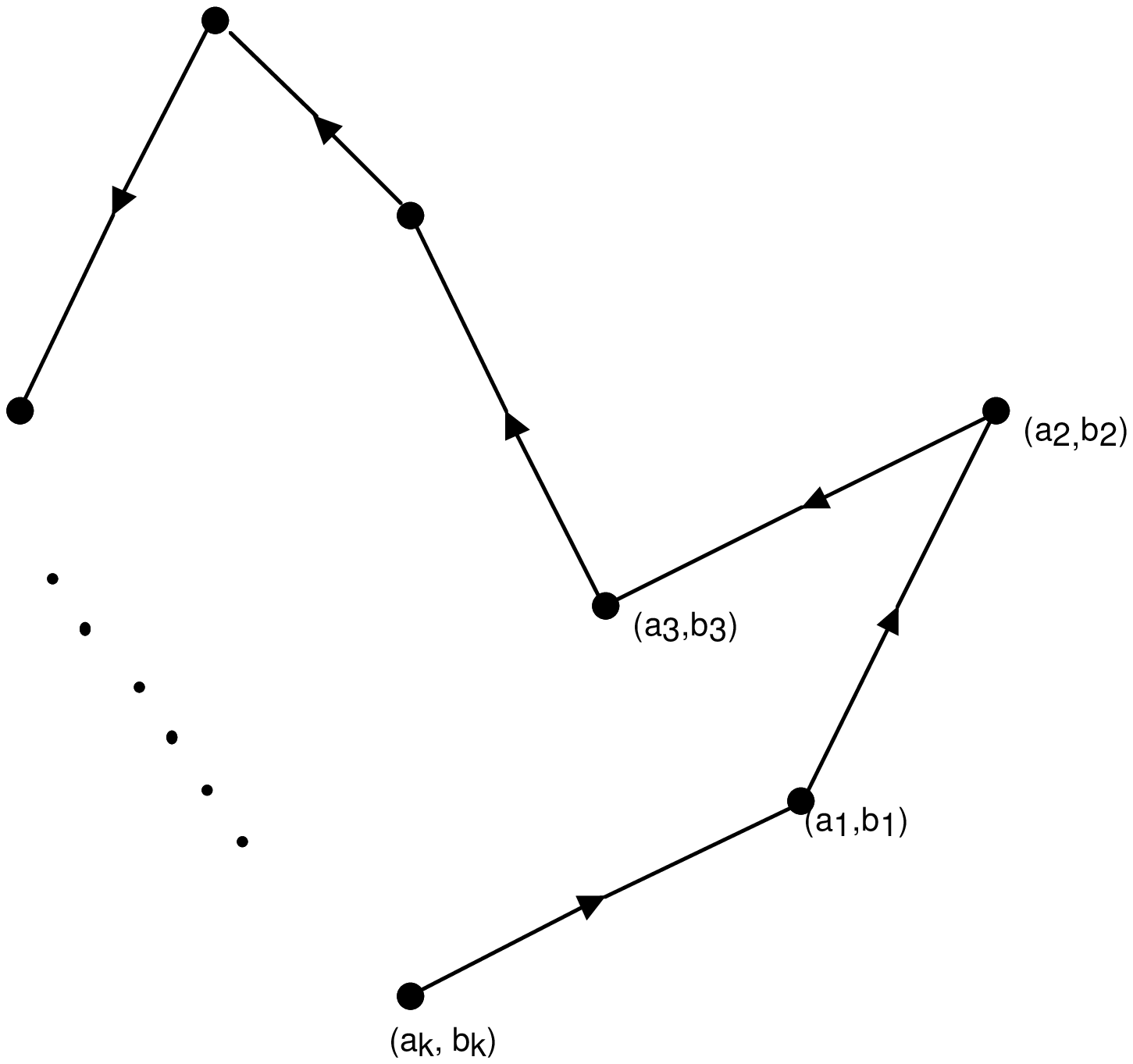} {3truein} 

We should 
remember to count with sign.  Orienting a polygon oppositely would change the sign of
$\int_{W.S.} B_{ij}$.  It is easily deduced that the result is 
\eqn\Bkphi{
\int_{W.S.} B_{ij}= \half B \sum _{i=2}^{k-1} \biggl|{\matrix{{a_{i+1} -a_i} &{a_i-a_1}\cr
	 {b_{i+1}-b_i}  & {b_i-b_1}\cr}\biggr|}
}
where $\Bigl|\matrix{a &b\cr c&d}\Bigr| = ad-bc$.  This means that the interaction eq.\kphi\ now becomes 
\eqn\intB{\eqalign{
\phi^{(k)}_{a_1b_1,a_kb_k} \phi^{(k-1)}_{a_kb_k,a_{k-1}b_{k-1}}
e^{i \half  B  \biggl| \matrix {{a_k -a_{k-1}} &{a_{k-1}-a_1}\cr
		      {b_k -b_{k-1}}  & {b_{k-1}-b_1} \cr} \biggr| } \ldots \ldots  &\cr                  
\phi^{(3)}_{a_4b_4,a_3b_3} 
 e^{i \half  B  \biggl| \matrix{{a_4 -a_3} &{a_3-a_1}\cr
		                  {b_4 -b_3}  & {b_3-b_1}\cr}  \biggr| }
\phi^{(2)}_{a_3b_3,a_2b_2}
 e^{i \half  B  \biggl| \matrix{{a_3 -a_2} &{a_2-a_1}\cr
		                  {b_3 -b_2}  & {b_2-b_1} \cr}  \biggr|}
\phi^{(1)}_{a_2b_2,a_1b_1}.&
}}
The reason for distributing the exponentials in this way will become clear in a moment. 
One could put an exponential between $\phi^{(k)}$ and $\phi^{(k-1)}$, but it would be 
identically 1, so we omitted it. We can introduce a notation which will make this look 
simpler. First note that the interactions always appear with a sum over indices.
\eqn\bigsum{
\sum_{a1b1,a_kb_k}{ \phi^{(k)}_{a_1b_1,a_kb_k} \phi^{(k-1)}_{a_kb_k,a_{k-1}b_{k-1}}
\ldots \phi^{(3)}_{a_4b_4,a_3b_3} \phi^{(2)}_{a_3b_3,a_2b_2} \phi^{(1)}_{a_2b_2,a_1b_1}}}
If we think of the fields as matrices this is just 
\eqn\trace{
Tr{ (\phi^{(k)}\cdot \phi^{(k-1)} \ldots \phi^{(2)} \cdot \phi^{(1)})}.
}
Now we define a product, called $*$, by
\eqn\star{
(\phi^{(2)}*\phi^{(1)})_{a_3b_3,a_1b_1}=
\sum_{a_2b_2}{e^ {\half B  \biggl| {\matrix{{a_3 -a_2} &{a_2-a_1}\cr
				{b_3-b_2}  & {b_2-b_1}\cr}} \biggr|}}  
	\phi^{(2)}_{a_3b_3,a_2b_2} \phi^{(1)}_{a_2b_2,a_1b_1}.
}	
Now the interaction with a $B$-field, eq.\intB, can be written
\eqn\startrace{
Tr(\phi^{(k)} *\phi^{(k-1)}* \ldots *\phi^{(2)} *\phi^{(1)}).
}
This is really nice.  It shows that to generalize the action eq\Obraneaction\ to include
a $B$-field we just need to replace ordinary matrix product with $*$-product.  For $B=0$
the $*$-product coincides with the ordinary product.  This point of view, that the fields take
value in another algebra, was of course the main point of \Connes.

In the case $B=0$ the fields with the constraints eq.\constraints\ and action can be 
rewritten to a $SYM$ theory on a dual $T^2$  \refs{\Wati,\WOS}.  Let us briefly repeat 
that construction for the case, $B\neq 0$.  Let us first express the basis of the Hilbert 
space in another form.  We want to think of the Hilbert space as $L_2$ functions on a dual torus
with radii, $1\over m_s^2 r_1$, $1\over m_s^2r_2$. Let the basis vector $|ab>$ correspond to 
$e^{iaxm_s^2r_1} e^{ibym_s^2r_2}$, then the operators $U_1$, $U_2$ become multiplication
operators
\eqn\Uonetwo{
	U_1=e^{ixm_s^2r_1}, \qquad U_2=e^{iym_s^2r_2}.
}
It is now easy to solve the constraints for the fields eq.\constraints
\eqn\solvecont{\eqalign{
X^1 &= -i2\pi{1 \over m_s^2} {\partial \over {\partial x}} + \sum_{a,b} X^1_{ab,00}
e^{iam_s^2r_1x}e^{ibm_s^2r_2y}\cr
X^2 &= -i2\pi {1\over m_s^2} {\partial \over {\partial y}} + \sum_{a,b} X^2_{ab,00}
e^{iam_s^2r_1x}e^{ibm_s^2r_2y}\cr
X^j & = \sum_{a,b} X^j_{ab,00} e^{iam_s^2r_1x} e^{ibm_s^2r_2y}\cr
\Psi & =\sum_{a,b}  \Psi_{ab,00} e^{iam_s^2r_1x} e^{ibm_s^2r_2y}.
}}
We see that $X^1$, $X^2$ become covariant derivatives and all other fields are multiplication operators.
These are exactly the fields of 2+1 dim $SYM$ on a torus of radii ${1\over m_s^2r_1}$, ${1\over m_s^2r_2}$.

All this is independent of $B$.  We saw that the only B-dependence was to change products of fields
to the $*$-product.  Let us see how the $*$-product looks in this basis.  We only need to consider the types
of operators which appear in the action.  We see from eq.\solvecont\ that these are the differential
operators, ${\partial\over{\partial x}}$, ${\partial\over{\partial y}}$, and multiplication operators.
${\partial \over{\partial x}}$ and $\partial \over{\partial y}$ are diagonal and it is easily seen from
eq.\star\ that for diagonal operators the $*$-product is equal to the usual product.  Let us now look at
two multiplication operators $\phi^{(1)}(x,y)$ and $\phi^{(2)}(x,y)$.  We have 
\eqn\fourieret{
\phi^{(i)}(x,y) = \sum_{a,b} \phi^{i}_{ab,00}e^{iam_s^2r_1x}e^{ibm_s^2r_2y}
}
with $\phi^{i}_{a_2b_2,a_2b_2} =  \phi^{i}_{(a_2-a_1)(b_2-b_1),00}$.
Plugging into eq.\star\ it is seen that 
\eqn\stjerne{
(\phi^{(2)}*\phi^{(1)})(x,y)= 
e^{-i\half {B \over m_s^4r_1r_2}( {\partial \over \partial x_2} {\partial \over \partial y_1}
- {\partial \over \partial y_2} {\partial \over \partial x_1})}
\phi^{(2)}(x_2,y_2) \phi^{(1)}(x_1,y_1)|_{x_2=x_1=x,y_2=y_1=y}.
}
Let us recapitulate what we have obtained so far. We consider one 0-brane 
on a $T^2$, $N_0=1$, and no membranes, $N_2=0$. The flux of the $B_{ij}$-field 
through the torus is $B$. We consider the limit coming from matrix theory. 
If $B=0$ the resulting theory is a 2+1 dim. SYM on a dual $T^2$. 
In terms of the M-theory variables the radii of the $T^2$ are
\eqn\param{\eqalign{
r_1^{'}= {1 \over m_s^{2}r_1}= {1 \over M^3RR_1} \cr 
r_2^{'}= {1 \over m_s^{2}r_2}= {1 \over M^3RR_2} 
}}
and the gauge coupling is 
\eqn\koblinggauge{
{1 \over g^2}= {m_s r_1 r_2 \over \lambda} = 
{R_1R_2 \over R}.
}
The gauge bundle 
on $T^2$ is trivial. This is a consequence of the fact that all the fields in 
eq.\solvecont\ are functions on $T^2$ instead of sections of a non-trivial bundle. 
Equivalently $c_1 = {1 \over 2\pi}\int trF = 0$. For any $B$ the only difference 
is that every time two fields are being multiplied in the action one should 
instead use the $*$-product.

When $B=0$ the $*$-product, of course, coincides with the usual product. 
Looking at eq.\star\ we see that the product has a periodicity in 
$B$ of $4\pi$. For $B=2\pi$ it is different from $B=0$. At first sight 
this is problematic, since the theory is known to have a periodicity in 
$B$ of $2\pi$. The puzzle is resolved by noting that there is a 
field redefinition which takes the theory at $B=2\pi$ into 
$B=0$. The field redefinition is 
\eqn\feltre{
\phi_{a_2b_2,a_1b_1} \rightarrow (-1)^{(a_2-a_1)(b_2-b_1)}
\phi_{a_2b_2,a_1b_1}.				
}
Thus the gauge theories actually have the correct $2\pi$ 
periodicity in $B$.

So far we have only discussed the case with $N_0=1$ and $N_2=0$, i.e. one 
0-brane and no 2-branes. The case with any $N_0$ goes in exactly 
the same way. It is now a $U(N_0)$ theory instead of a $U(1)$ 
theory. Nothing else is changed. Especially the same form of the 
$*$-product should be used, except that now the fields are 
$N_0 \times N_0$ matrices.

%%% ------------------------- CUT HERE ---------------------------------%
% ===================================================================== %
% Non-trivial gauge bundles
% ===================================================================== %
\newsec{Non-trivial gauge bundles}

In the previous section we only considered cases with no membranes, 
$N_2 =0$. What about $N_2 \neq 0$? In the case $B =0$ we know 
the answer. Here the final theory is obtained by T-duality on both 
circles. After T-duality we get the decoupled theory of 
$N_0$ D2-branes with $N_2$ D0-branes dispersed in the 2-branes. 
0-branes in 2-branes is just magnetic flux. In other words now 
it is a $U(N_0)$ theory with a non-trivial bundle on $T^2$. The 
first Chern class is $c_1= {1 \over 2\pi}\int trF = N_2 $. In the 
previous section we saw that for $B \neq 0$ and $N_2=0$ the theory 
became a $U(N_0)$ theory with $c_1=0$ and deformed by the 
$*$-product. All these theories have radii and coupling given by 
eq.\param. The obvious guess is now that the case with $B \neq 0$
and $N_2 \neq 0$ was described by a $U(N_0)$ theory with 
$c_1 = N_2$ and an action deformed by the $*$-product. However this 
cannot be true, at least not in this naive sense. The reason is 
that if the bundle is non-trivial we really need to define the 
fields in coordinate patches. The $*$-product does not transform 
correctly under change of patch. To make it do so we would have 
to replace $\partial \over \partial x$, $\partial \over \partial y$ 
by covariant derivatives. Even if this is possible there are other 
reasons to doubt that this is correct. Firstly $B \rightarrow 
B + 2\pi$ is not a symmetry in the presence of 2-branes. It is 
only a symmetry if one changes the number of 0-branes: 
$N_0 \rightarrow N_0 - N_2$. For $N_2=1$ one could always 
change to $N_0 =0$. If the above guess was correct this would 
lead to strange connections between ``$U(0)$'' and $U(N)$
theories. 

Another reason to doubt this naive guess is the following. 
When one uses Sen's and Seiberg's prescription \refs{\Sen,\Seiwhy} to derive 
a matrix model the energy has the form
\eqn\matenergi{
E = \sqrt{({N \over R})^2 + P_{\perp}^2 + m^2} = {N \over R} 
+ \half {R \over N}(P_{\perp}^2 + m^2) + \ldots .
}
Here it is written for uncompactified M-theory, but a similar 
expression is valid for all compactifications. The point is 
that in the limit $R \rightarrow 0$, the first term goes to 
infinity, the second term stays finite(after rescaling of all 
energies) and the terms indicated by dots vanish. The first 
term goes to infinity but is fixed and independent of 
any dynamics. Therefore we can ignore it and just keep the 
second term. A matrix theory Hamiltonian always gives the 
second term. When we change N the theory changes drastically. 
For instance the gauge group changes. In other words when the 
infinite term is changed we expect the finite term to change 
drastically. Let us now look at our situation. With $N_0$ 
0-branes, $N_2$ 2-branes and a $B_{ij}$-field flux $B$. Here 
the energy takes the form
\eqn\energib{
E = {N_0 + BN_2 \over R} + finite.
}
For $B \neq 0$ the infinite term changes when $N_2$ is changed. 
So following the remarks above we expect the theory to 
change drastically. Specifically it is probably not enough 
to change the bundle, but also radii and gauge coupling 
changes. 

Whether or not the case of $N_2 \neq 0$ is solved by just 
changing the first Chern class, there is another way of 
solving it. This is the subject of next section.

%%% ------------------------- CUT HERE ---------------------------------%

% ===================================================================== %
% Incorporating 2-branes
% ===================================================================== %
\newsec{Incorporating 2-branes}

In this section we will obtain the matrix model for the general case, 
with $N_0$ 0-branes,$N_2$ 2-branes and a flux $B$. We will do that 
by performing a T-duality to transform to the case $N_2=0$

For a review of T-duality, see \Giveon. The T-duality group for 
IIA on $T^2$ contains an $SL(2,Z)$ subgroup which acts as follows. 
It leaves the complex structure of $T^2$ invariant. Define a 
complex number in the upper halfplane, $\rho = B +iV$. Here V is 
the volume of the torus measured in string units and $B$ is the 
flux of $B_{ij}$ through the torus. In our case $V=m_s^2 r_1r_2$. An 
element $\pmatrix{a&b \cr c&d \cr} \in SL(2,Z)$ acts as follows
\eqn\slaction{\eqalign{
\rho^{'}&= {a\rho +b \over c\rho +d} \cr
\pmatrix{N_0^{'} \cr -N_2^{'}\cr} &= \pmatrix{a&b \cr c&d \cr} 
\pmatrix{N_0 \cr -N_2 \cr}.
}}
Let us use this in our case. Let $Q$ be the greatest common 
divisor of $N_0$ and $N_2$. Write 
\eqn\stoerste{\eqalign{
N_2 &= Q \tilde{N_2} \cr
N_0 &= Q \tilde{N_0}.
}}
Since $\tilde N_0$,$\tilde N_2$ are relatively prime we can choose 
$a,b$ such that $a \tilde{N_0} - b \tilde{N_2}=1$. Let us now 
perform a T-duality transformation with the matrix 
\eqn\tmatrix{
\pmatrix{a&b \cr \tilde N_2 & \tilde N_0 \cr }.
}
An easy calculation gives the new radii, $B_{ij}$ flux, 0-brane and 
2-brane numbers
\eqn\nyparam{\eqalign{
r_1^{'} &= {r_1 \over \tilde N_0 + \tilde N_2 B} \cr
r_2^{'} &= {r_2 \over \tilde N_0 + \tilde N_2 B} \cr
B^{'} &= {aB+b \over \tilde N_0 + \tilde N_2 B} \cr
N_0^{'} &=Q \cr
N_2^{'} &=0. 
}}
The string mass is invariant
\eqn\strengmasse{
m_s^{'} = m_s 
}
and the new coupling is 
\eqn\nykobling{
\lambda^{'}= {\lambda \over \tilde N_0 + \tilde N_2 B}.
}
In these formulas we have taken the matrix theory limit in the 
quantities which have a non-zero limit. We remark that the 
denominator $\tilde N_0 + \tilde N_2 B$ is positive since 
\eqn\naevner{
P_{-} = {\tilde N_0 + \tilde N_2 B \over R}
}
and $P_-$ is positive as always in matrix theory. We now see 
that the parameters of the theory go to zero and infinity 
in exactly the same way as in last section. This means that 
we are in exactly the same situation as in last section.

In other words the matrix theory is a 2+1 dim. SYM on a $T^2$ 
with gauge group $U(Q)$ where $Q$ is the greatest common divisor of 
$N_0$ and $N_2$. The action is deformed with the $*$-product 
with a value of $B$ equal to 
\eqn\vaerdi{
B^{'} = {aB+b \over \tilde N_0 + \tilde N_2 B}.
}
The $T^2$ has radii
\eqn\stor{\eqalign{
r_1^{''}= {1 \over m_s^{'2}r_1^{'}}= {\tilde N_0 + \tilde N_2 B \over M^3RR_1} \cr 
r_2^{''}= {1 \over m_s^{'2}r_2^{'}}= {\tilde N_0 + \tilde N_2 B \over M^3RR_2} 
}}
and the gauge coupling is 
\eqn\kobgauge{
{1 \over g^2}= {m_s^{'}r_1^{'}r_2^{'} \over \lambda^{'}} = 
{R_1R_2 \over R(\tilde N_0 + \tilde N_2 B)}.
}
The SL(2,Z) duality employed has a very easy geometric 
interpretation if one performs a T-duality on a single 
circle as in \DougHull. Here $N_0,N_2$ parametrize which
homology cycle the D-strings wrap. The factor $\tilde N_0 + \tilde N_2 B$ 
is just the length of the D-string. The T-duality transformation is just 
a geometric change of $\tau$-parameter of the torus.

%%% ------------------------- CUT HERE ---------------------------------%

% ===================================================================== %
% Conclusion
% ===================================================================== %
\newsec{Conclusion}

 It was explained in \refs{\Wati,\WOS} how to describe 0-branes on $T^2$ 
by working on the covering space ${\bf R}^2$ and modding out by 
translations. We did this in the presence of a background 
B-field. This enabled us to get a matrix theory of M-theory on 
$T^2$ with a background $C_{-12}$. The result agrees with 
\refs{\Connes, \DougHull} and is a gauge theory on a noncommutative 
torus. 
	
There are some interesting aspects of this. In the case $B=0$ this 
procedure leads to a 2+1 dim SYM which is exactly the same as the 
theory on the D2-brane in the T-dual picture. In other words the 
procedure of compactifying the 0-branes agrees with 
T-duality. For $B \neq 0$ this is not the case. T-duality 
 does not give a theory on a finite torus when $B \neq 0$. 
This is the whole reason for all this interest in $B \neq 0$. 
This means that working with 0-branes on the covering space is 
not the same as T-duality. We believe, of course, that T-duality 
still is true. The point is just that the T-dual description 
is not simpler. The T-dual description is the theory on D2-branes 
wrapped on a dual $T^2$ which is again shrinking. To extract a 
well defined action out of that one has to expand the full 
Born-Infeld action as advocated in {\Li}. It would indeed 
be interesting to use the noncommutative theory to put 
constraints on the full Born-Infeld action. All the higher 
derivative terms should come out of this.

Originally it was thought that compactifications of 
M-theory could be gotten by compactifying the 
0-brane quantum mechanics. That was indeed the case 
for toroidal compactifications up to $T^3$. For other 
compactifications certain degrees of freedom were missing. 
It was later realized that the correct way of obtaining 
the matrix model was to use string dualities in order 
to realise the theory as a theory living on a brane decoupled 
from gravity. In the case of $C_{-12} \neq 0$ we are in some 
sense back to the first philosophy. We can obtain the 
final theory starting with the 0-brane theory but 
we do not know how to realise it as a sensible limit of 
a theory on a brane.

It is an interesting question whether these new theories 
make sense as renormalizable quantum theories. In the case 
of $B =0$ we know that the procedure of putting 0-branes 
on the covering space gives a renormalizable theory 
up to $T^3$ and not for higher tori. So certainly arguing 
that this procedure should give a well defined theory 
is wrong. However, one might hope that the question of 
renormalisability is related to the ``number of degrees of 
freedom''. In that sense the theory on $T^d$ with $B \neq 0$ 
behaves like the theory on $T^d$ with $B=0$ and we might 
expect that the noncommutative theories are well defined up 
to $T^3$. Realizing these theories as theories on branes 
would resolve this issue, but as discussed above this 
might require knowing the full Born-Infeld action.

It will be very interesting to see what the methods 
of noncommutative geometry can teach us about string 
theory and the other way round.

%%% ------------------------- CUT HERE ---------------------------------%
%=======================================================================%
% Acknowledgments
%=======================================================================%
\bigbreak\bigskip\bigskip
\centerline{\bf Acknowledgments}\nobreak
We thank Ori Ganor for useful discussions and Sangmin Lee for careful reading
the manuscript. The research of Morten  Krogh was supported by the Danish 
Research Academy.

%%% ------------------------- CUT HERE ---------------------------------%

\listrefs
\bye
\end